\documentclass[%
 reprint,
nofootinbib,
 amsmath,amssymb,
 aps,
 prl,
]{revtex4-2}
\usepackage{graphicx}% Include figure files
\usepackage{dcolumn}% Align table columns on decimal point
\usepackage{bm}% bold math
\usepackage{hyperref}% add hypertext capabilities
\usepackage{color}
\usepackage{ragged2e}
\usepackage{subfig}
\usepackage{tabularx}
\usepackage{enumitem}

\newcommand{\mc}{\mathcal}

\begin{document}

\preprint{APS/123-QED}

\title{Testing spooky action between free-traveling electron-positron pairs}

%\title{Quantum entanglement in free-traveling electron-positron pairs}

\author{Leyun \surname{Gao}}
\email[]{seeson@pku.edu.cn}
\affiliation{State Key Laboratory of Nuclear Physics and Technology,\\School of Physics, Peking University, Beijing, 100871, China}

\author{Alim \surname{Ruzi}}
\email[]{alim.ruzi@pku.edu.cn}
\affiliation{State Key Laboratory of Nuclear Physics and Technology,\\School of Physics, Peking University, Beijing, 100871, China}

\author{Qite \surname{Li}}
\email[]{liqt@pku.edu.cn}
\affiliation{State Key Laboratory of Nuclear Physics and Technology,\\School of Physics, Peking University, Beijing, 100871, China}

\author{Chen \surname{Zhou}}
\email[]{czhouphy@pku.edu.cn}
\affiliation{State Key Laboratory of Nuclear Physics and Technology,\\School of Physics, Peking University, Beijing, 100871, China}

\author{Qiang \surname{Li}}
\email[]{qliphy0@pku.edu.cn}
\affiliation{State Key Laboratory of Nuclear Physics and Technology,\\School of Physics, Peking University, Beijing, 100871, China}

%\date{\today}% It is always \today, today,
%             %  but any date may be explicitly specified

\begin{abstract}
Quantum entanglement is a cornerstone of quantum mechanics. While the entanglement of confined electron pairs has been established early on, the entanglement of free-traveling electron pairs, particularly at high energies, remains largely unexplored due to the substantial challenges involved in measuring the spins of free-traveling electrons. In this study, we investigate the entanglement and the Bell inequality violation of free-traveling electron-positron pairs generated in a fixed-target experiment. This experimental setup facilitates the creation of a controllable source of entangled electron-positron pairs, where entangled events are produced in specific phase spaces. Based on this source and the prior knowledge of the entangled state, we demonstrate the feasibility of measuring the polarization correlations of the entangled $e^+e^-$ pairs through their individual secondary scatterings off two separate additional targets.
\end{abstract}

%\keywords{Suggested keywords}%Use showkeys class option if keyword
                              %display desired
\maketitle

%\tableofcontents

\section{Introduction}

In quantum mechanics, many-body systems often exhibit non-classical correlations, where the state of each subsystem is interdependent with those of the others, regardless of their spatial separation. This phenomenon, known as quantum entanglement~\cite{Einstein:1935rr}, is fundamental to the unique capabilities of quantum science and technology. The most distinguishing feature of quantum entanglement is the violation of Bell inequalities~\cite{Bell:1964kc,Freedman:1972zza,Clauser:1978ng}, which demonstrates the non-locality of our universe.

Studies of quantum entanglement in high-energy physics test quantum theory at the highest energy frontier~\cite{Fabbrichesi:2021npl,Barr:2021zcp,Ashby-Pickering:2022umy,Fabbrichesi:2023cev,Barr:2024djo} and provide a means to constrain free parameters in various physical models through experimental investigations~\cite{Fabbrichesi:2022ovb,Fabbrichesi:2024xtq,LoChiatto:2024dmx}. The ATLAS Collaboration has recently observed quantum entanglement involving top quarks at a center-of-mass energy of 13 TeV, marking the highest energy measurement of quantum entanglement to date~\cite{ATLAS:2023fsd}.

As early as 1950, Chien-Shiung Wu et al. published a paper~\cite{Wu:1950zz} in which the angular correlation of two Compton-scattered photons arising from electron-positron annihilation are measured. This work can be considered the first experiment on quantum entanglement~\cite{Yang:2015txa}. The experimental setup can be extended to measure correlations between free-traveling charged leptons, as will be discussed in this article.

Charged leptons in the Standard Model serve as qubits in quantum information theory, with their production, scattering, and decay processes precisely described by quantum electrodynamics. Most studies on charged lepton quantum entanglement have concentrated on the decaying tau leptons~\cite{Babson:1982ms,Privitera:1991nz,Dreiner:1992gt,Ehataht:2023zzt,Ma:2023yvd,Altakach:2022ywa,Fabbrichesi:2024wcd}, while less attention has been given to electrons and muons.
%Due to the symmetry-breaking nature of the electroweak interaction, the spins of tau leptons can be indirectly measured through the distributions of their decay products. Based on this principle, testing Bell inequalities via the $e^+e^- \to \tau^+\tau^-$ process has been investigated through simulations and proposed for experiments at LEP from the outset~\cite{Babson:1982ms,Privitera:1991nz,Dreiner:1992gt}. Although a violation has not yet been observed, numerous future proposals aim to conduct these tests through various $\tau^+\tau^-$ processes at current and upcoming colliders, such as Belle II~\cite{Ehataht:2023zzt}, CEPC~\cite{Ma:2023yvd}, ILC~\cite{Altakach:2022ywa}, and FCC-ee~\cite{Altakach:2022ywa,Fabbrichesi:2024wcd}.
%In a lepton on-target experiment, the accessible incident lepton beams typically include muon beams at scales of up to 100 GeV and electron/positron beams at scales of up to 10 GeV~\cite{Pilato:2022wvg,Aniculaesei:2022kwl}. The center-of-mass energies are usually much lower than those found in typical lepton colliders, which prevents the production of $\tau^+\tau^-$ pairs through $e^+e^-$ annihilation and necessitates a high minimum incident positron beam energy of approximately 44 GeV for the production of $\mu^+\mu^-$ pairs.
In a previous study~\cite{Gao:2024leu}, we investigated the long-neglected potential of muons for quantum state tomography and Bell inequality tests, specifically in the context of $\mu^-e^-$ scattering in muon on-target experiments, utilizing a kinematic approach~\cite{PhysRevD.107.116007,Cheng:2024rxi}. In this article, we examine the case of Bhabha scattering within a similar theoretical framework. This leads to a new method for producing high-energy entangled electron-positron pairs. 
%Through this process, there is no lower limit on the center-of-mass energy imposed by the final state, unlike in the cases of $\mu^+\mu^-$ and $\tau^+\tau^-$ final states. Additionally, both the entanglement strength and the measurement precision of the scattering angles can benefit more from a low-energy $e^+e^-$ final state compared to the $\mu^-e^-$ final state, as will be demonstrated numerically in the following sections.
Both the entanglement strength and the scattering angles can benefit more from the $e^+e^-$ final state compared to the $\mu^-e^-$ final state, as will be demonstrated numerically in the following sections.

Various experimental measurements for probing entanglement and the violation of the Bell inequality can be an eager with the entangled pair source established. Successful examples involving fixed electrons were conducted two decades ago with electron pairs confined in semiconductor quantum dots~\cite{Petta:2005dew}. In those experiments, spin-entangled two-electron states were prepared, coherently manipulated, and measured, facilitating quantum computation within a solid-state system. However, in high energy experiments, measuring the spin of a single traveling electron poses a significant challenge due to interference from its orbital motion~\cite{doi:10.1080/00107510110102119}. Instead, it is feasible to measure the average polarization of a beam of electrons or positrons by analyzing the distributions of the beam's scattering products following polarization-sensitive processes such as Mott scattering, Møller scattering, and Compton scattering~\cite{Gaskell2023}. Utilizing these strategies, the spin correlation of electron-positron pairs generated by positron targeting can be measured through two individual polarized secondary scattering processes: Bhabha scattering for positrons and Møller scattering for electrons. Simulation results and feasibility considerations of this proposal are presented in this article following the discussion of the entangled pair source.

\section{Theoretical framework}

After Bhabha scattering between the incoming positron and the target electron, the density matrix of the outgoing particles can be expressed as a function of kinematic quantities, specifically the scattering angles of the final state particles. Assuming that the initial state is unpolarized, the final state density matrix $\rho_f$ can be derived as follows:
\begin{equation}
\rho_{s^\prime_3s^\prime_4s^{\phantom{\prime}}_3s^{\phantom{\prime}}_4} = \frac{1}{4}\sum_{s^{\phantom{\prime}}_1s^{\phantom{\prime}}_2}\left(\frac{\mathcal{M}_{s^\prime_3s^\prime_4s^{\phantom{\prime}}_1s^{\phantom{\prime}}_2}\mathcal{M}^*_{s^{\phantom{\prime}}_3s^{\phantom{\prime}}_4s^{\phantom{\prime}}_1s^{\phantom{\prime}}_2}}{\sum_{s^{\prime\prime}_3s^{\prime\prime}_4}\left|\mathcal{M}_{s^{\prime\prime}_3s^{\prime\prime}_4s^{\phantom{\prime}}_1s^{\phantom{\prime}}_2}\right|^2}\right).
\end{equation}
Here, $s_1$ and $s_2$ represent the helicities of the positron and electron before scattering, while $s_3$ and $s_4$ denote the helicities after scattering. The polarized scattering amplitude, $\mathcal{M}$, is calculated at the tree level as the coherent superposition of the s- and t-channel contributions.

Entanglement can be quantified by several measures~\cite{Fedida:2022izl}, which are known as a class of non-negative functions called entanglement monotones~\cite{Horodecki:2009zz,Chitambar:2018rnj}. One such measure is \emph{concurrence}~\cite{Wootters:1997id,Gingrich:2002ota,Hill:1997pfa}. For a mixed state of two qubits, the concurrence is defined as
\begin{equation}
\mc{C}(\rho_f) = \max\{0, \lambda_1 - \lambda_2 - \lambda_3 - \lambda_4\},
\end{equation}
where $\lambda_i$ ($\lambda_i \geq \lambda_j,\ \forall i < j$) are the square roots of the eigenvalues of the matrix $\rho_f(\sigma_2 \otimes \sigma_2)\rho_f^*(\sigma_2 \otimes \sigma_2)$. If $\mc{C} > 0$, the two-qubit system is entangled. While $\mathcal{C}(\rho_f)$ is generally reference-frame-dependent~\cite{Gingrich:2002ota}, $\mathcal{C}(\rho_f)$ here inherits Lorentz invariance from the scattering amplitude $\mathcal{M}$.

In addition, we can test the violation of the \emph{CHSH inequality}, $I_2 \leq 2$ \cite{Clauser:1969ny}, which is the Bell inequality for a two-qubit system, by evaluating the optimal (maximal) $I_2$ \cite{Horodecki:1995nsk} as
\begin{equation}
I_2 = 2\sqrt{\lambda_1 + \lambda_2},
\end{equation}
where $\lambda_1$ and $\lambda_2$ are the two largest eigenvalues of the matrix $C^\mathrm{T}C$, and $C$ is the correlation matrix calculated by $C_{ij} = \text{Tr}\left(\rho_f\left(\sigma_i \otimes \sigma_j\right)\right)$. Any value of $I_2$ exceeding 2 violates the CHSH form of Bell inequalities and strongly rejects the non-entanglement implications of local hidden variable theories.

\section{Establishment of a controllable entangled electron-positron source}

\begin{figure}
\centering
\subfloat[polar angles]{\label{fig:theta}\includegraphics[width=.5\linewidth]{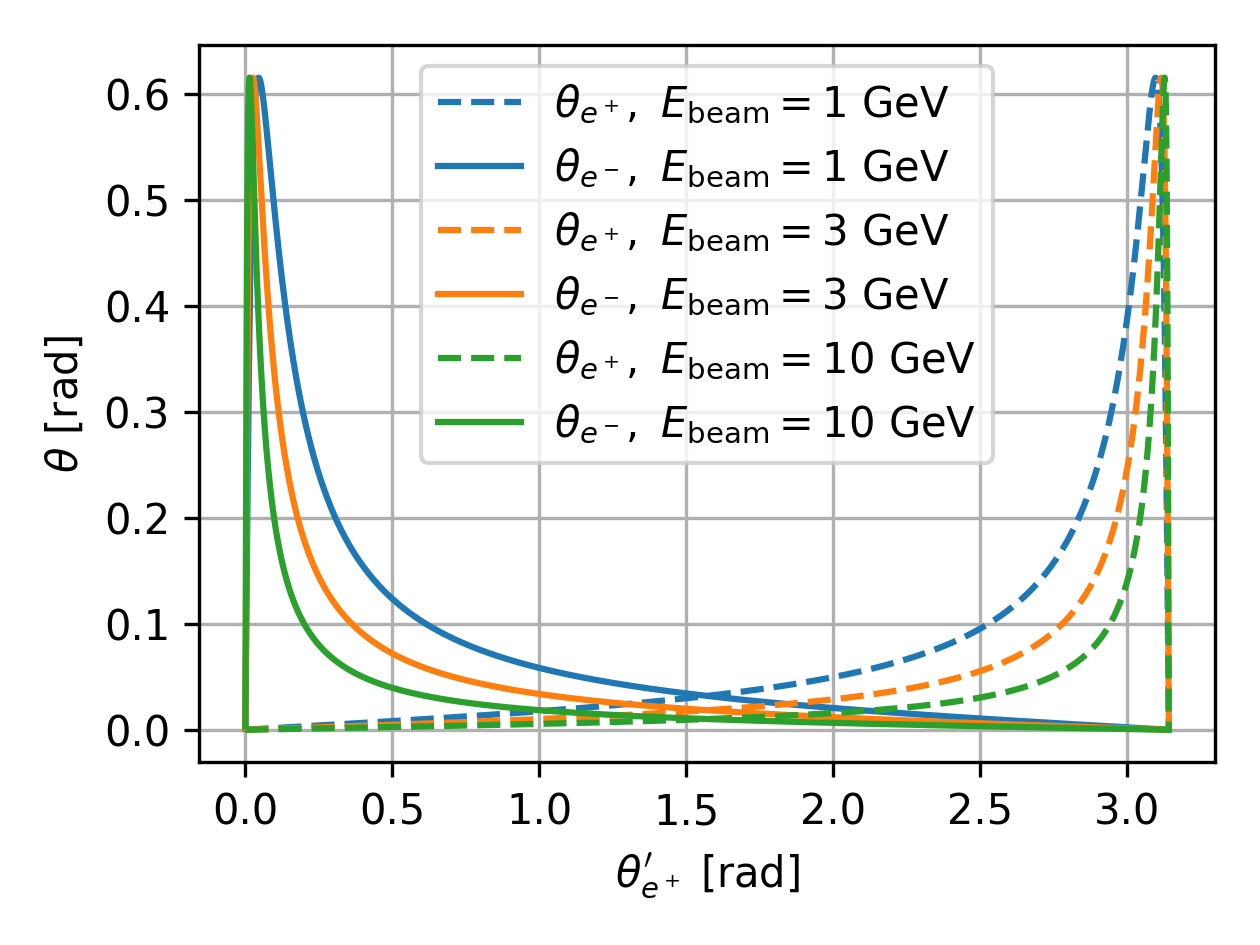}}
\subfloat[energies]{\label{fig:E}\includegraphics[width=.5\linewidth]{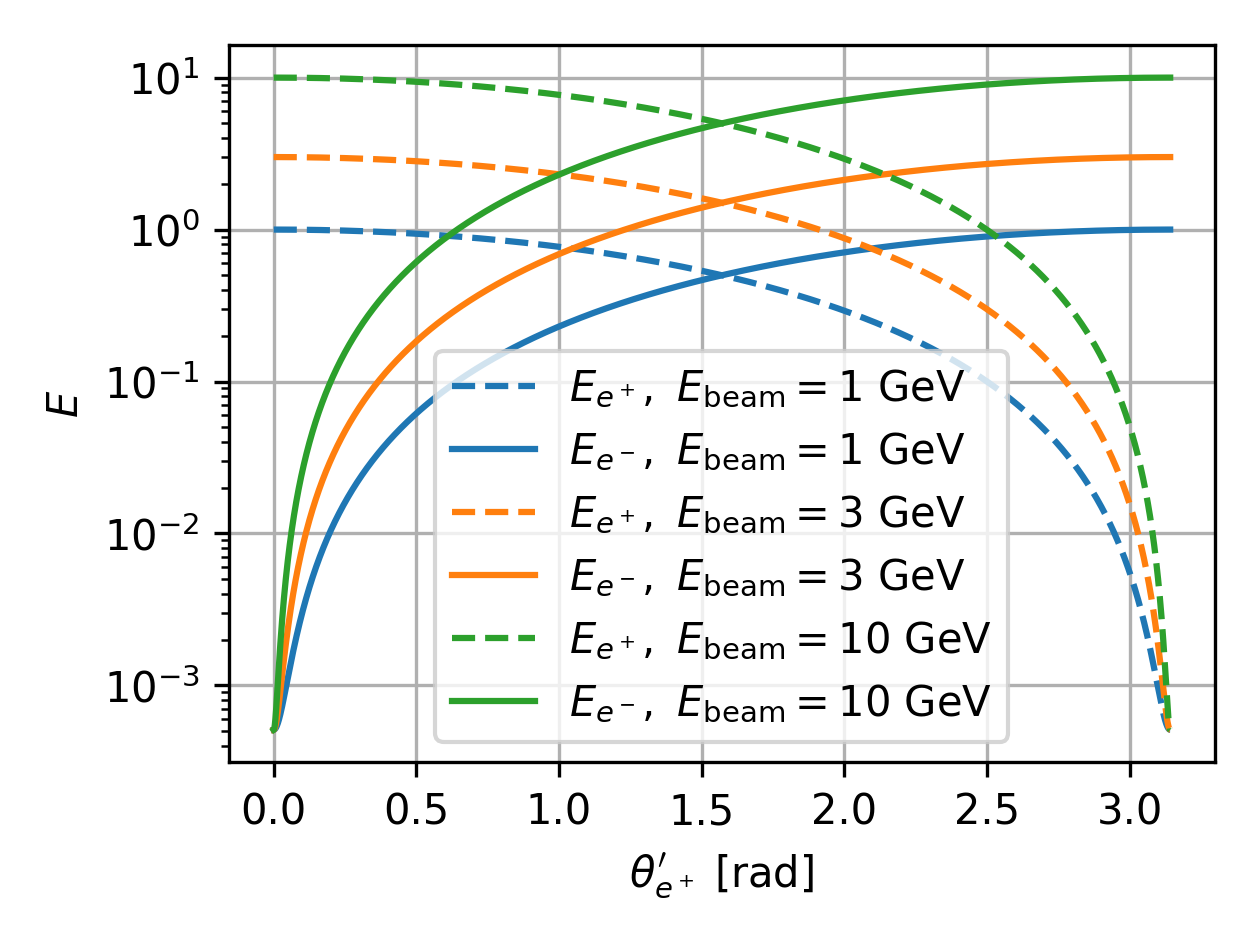}}
\caption{\justifying (a) Polar angles $\theta_{e^-}$ and $\theta_{e^+}$, and (b) energies $E_{e^-}$ and $E_{e^+}$ of the final state electron and positron in the lab frame, as functions of the polar angle of the final state positron in the center-of-mass frame, $\theta'_{e^+}$.}
\label{fig:theta-E}
\end{figure}

We use MadGraph5\_aMC@NLO 3.5.5~\cite{Alwall:2014hca} to simulate Bhabha scattering at tree-level QED, incorporating lepton masses. We consider positron beams with energies of 1, 3, and 10 GeV. Let $\theta_{e^-}$ ($\theta'_{e^-}$) and $\theta_{e^+}$ ($\theta'_{e^+}$) denote the polar angles of the final state electron and positron momenta in the lab (center-of-mass) frame, respectively. Within the studied energy range, spin entanglement is observed with a lower bound of approximately 1.3 rad for $\theta'_\mu$, in agreement with Refs.~\cite{Fedida:2022izl,Cervera-Lierta:2017tdt}. Due to the symmetry illustrated in Fig.~\ref{fig:theta} and the limitations of the event generator, events with $\theta'_{e^+} > 3$ can only be generated alongside $\theta'_{e^+} < \pi - 3$, which has a significantly larger cross section, rendering the generation of the former inefficient. In this study, we focus solely on the range $\theta'_{e^+} \leq 3$ and applies the corresponding final state lepton pseudorapidity ($\eta$, where $\eta = -\ln[\tan(\theta/2)]$) selections in event generation.

\begin{figure*}
\centering
%\subfloat[]
{\includegraphics[width=.32\linewidth]{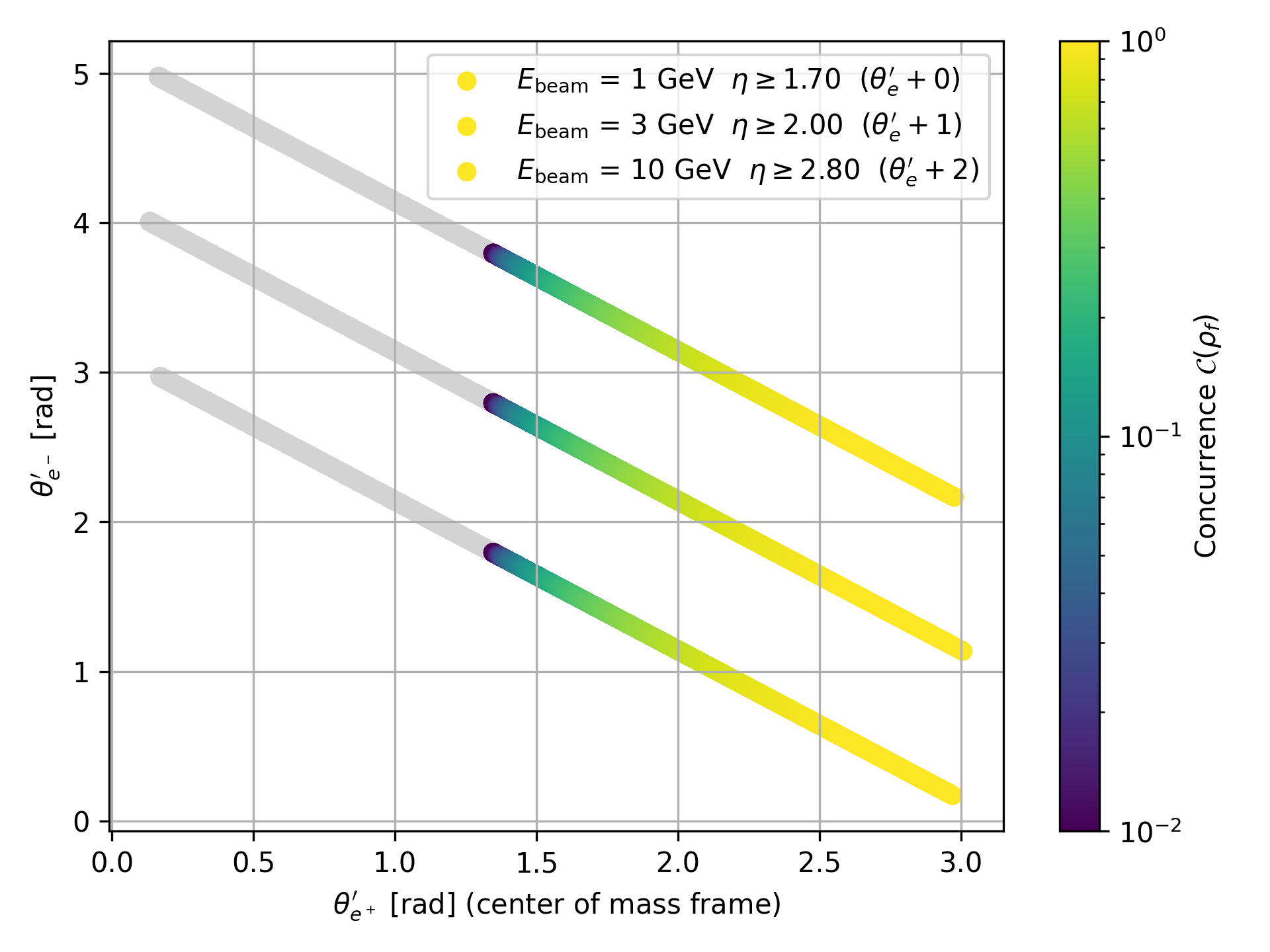}}
%\subfloat[]
{\includegraphics[width=.32\linewidth]{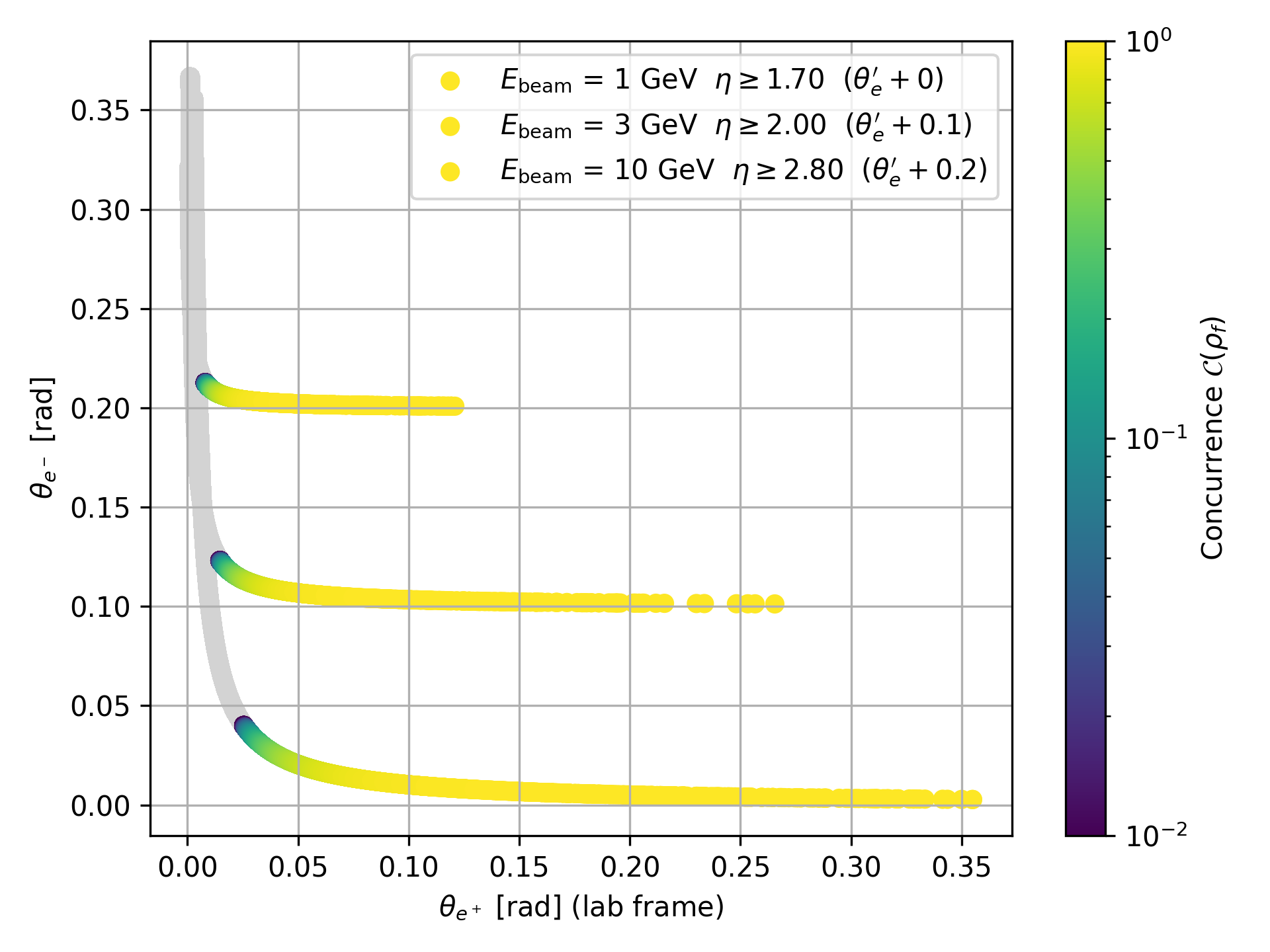}}

%\subfloat[]
{\includegraphics[width=.32\linewidth]{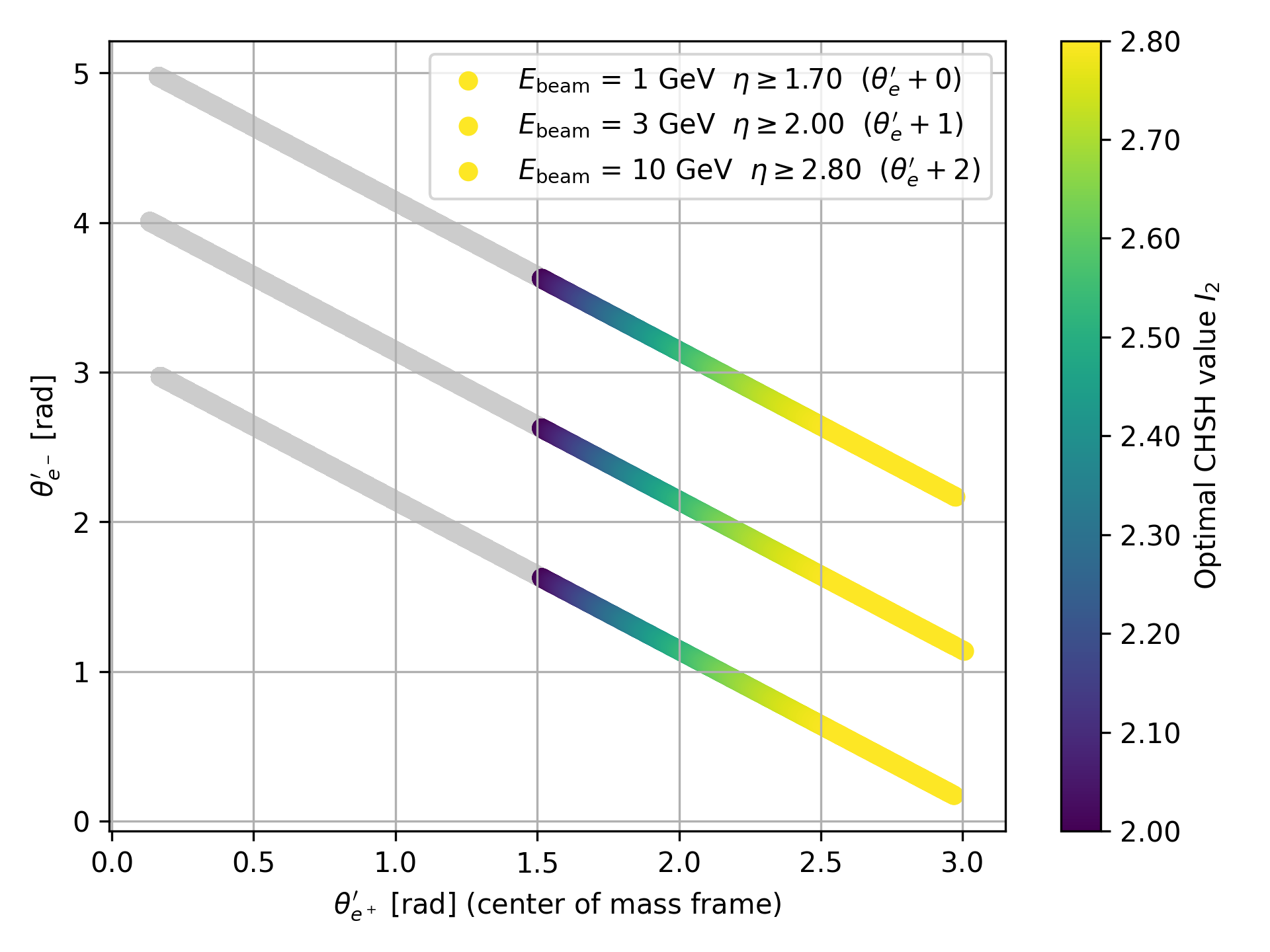}}
%\subfloat[]
{\includegraphics[width=.32\linewidth]{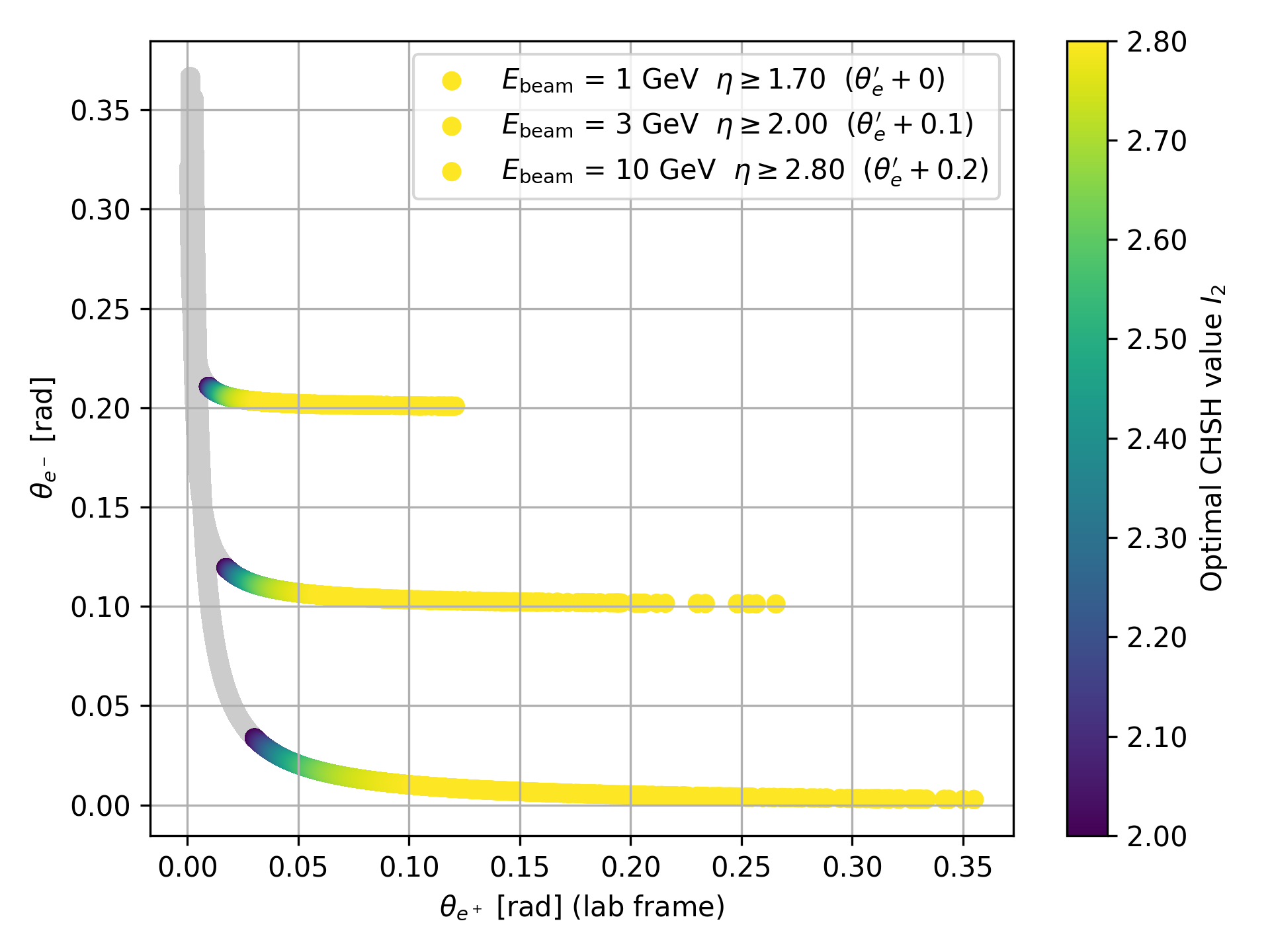}}

\caption{\justifying Final state scatter plots colored by (top) the concurrence $\mathcal{C}(\rho_f)$ and (bottom) the optimal CHSH value $I_2$. These plots correspond to various incoming positron energies in (left) the center-of-mass frame and (right) the lab frame. The light gray regions indicate where $\mathcal{C}(\rho_f) = 0$ or $I_2 \leq 2$. Events are generated with the labeled minimum pseudorapidity requirements.}
\label{fig:qe}
\end{figure*}

\begin{table*}[t]
\caption{\justifying Results on quantum entanglement and the violation of the CHSH inequality, along with the final state kinematic information, for the three studied positron beam energies.}
\label{tab:summary}
\setlength{\tabcolsep}{1em}
\begin{tabularx}{\linewidth}{rrrrrrrrr}
\hline\hline
$E_\mathrm{beam}$/GeV &
$E_\mathrm{COM}$/GeV &
$\mc{C}^{\max}(\rho_f)$ &
$I_2^{\max}$ &
$E_{e^+}^{\min}$/GeV &
$E_{e^-}^{\min}$/GeV &
$\theta_{e^+}^{\min}$/rad &
$\theta_{e^-}^{\min}$/rad &
$\sigma_{\mathrm E}$/\textmu b \\
\hline
 1 & 0.032 & 0.9996 & 2.8281 & 0.008 & 0.389 & 0.0255 & 0.0028 & 243.6 \\ % 0.00700 * 3.48e+10
 3 & 0.055 & 0.9997 & 2.8282 & 0.023 & 1.166 & 0.0147 & 0.0016 &  82.1 \\ % 0.00429 * 1.914e+10
10 & 0.101 & 0.9997 & 2.8282 & 0.074 & 3.890 & 0.0081 & 0.0009 &  26.5 \\ % 0.00706 * 3.749e+09
\hline\hline
\end{tabularx}
\end{table*}

The scatter plots in Fig.~\ref{fig:qe} illustrate the angular distributions of concurrence $\mathcal{C}(\rho_f)$ and the optimal CHSH value $I_2$ in both the center-of-mass and lab frames. In comparison to the results for $\mu^-e^- \to \mu^-e^-$ presented in Ref.~\cite{Gao:2024leu}, the angular ranges exhibiting entanglement witness $\mathcal{C}(\rho_f) > 0$ in the center-of-mass frame are significantly broader. The theoretical upper limits for both $\mathcal{C}(\rho_f)$ and $I_2$ in quantum mechanics are nearly reached as $\theta'_{e^+}$ approaches 3. For all three energies studied, over 70\% of the events with $\mathcal{C}(\rho_f) > 0$ display $I_2 > 2$, thus violating the CHSH inequality. Further details are summarized in Tab.~\ref{tab:summary}, where $E_{e^-}^{\min}$ and $\theta_{e^+}^{\min}$ correspond to the dark blue ends of the $\mathcal{C}(\rho_f)$ plots in Fig.~\ref{fig:qe}, indicating the starting points for $\mathcal{C}(\rho_f) > 0$. $E_{e^+}^{\min}$ and $\theta_{e^-}^{\min}$ correspond to the yellow ends, constrained by $\theta'_{e^+} \leq 3$.

To analyze the expected yields of entangled events in future positron on-target experiments, we define the entangled cross section $\sigma_{\mathrm{E}}$ as the scattering cross section $\sigma$ multiplied by the ratio of events with $\mathcal{C}(\rho_f) > 0$. While entanglement can generally be measured across a broad energy range, electron-positron pairs produced under the 1 GeV configuration exhibit the most significant angular separation, making them particularly suitable for subsequent polarization correlation measurements. This configuration also yields a high production rate of entangled events. Assuming a 1 GeV positron beam with a flux of $10^{12}/\mathrm{s}$ directed at a $10\ \mathrm{cm}$ thick aluminum target, the expected entangled event rate is $1.9 \times 10^9$/s.  % 243.6e-6 * 1e-24 * 10 * 2.7 / 27 * 13 * 6.022e23 * 1e12
Additionally, consider the region where $0.05\ \mathrm{rad} \leq \theta_{e^+} \leq 0.1\ \mathrm{rad}$, which accounts for 23.4\% of all events with $\mathcal{C}(\rho_f) > 0$. This region ensures $E \geq 0.094\ \mathrm{GeV}$ and $\theta \geq 0.0103\ \mathrm{rad}$, as shown in Fig.~\ref{fig:theta-E}. It provides excellent measurability and electron-positron separability. It also indicates strong entanglement and a violation of the CHSH inequality, with $\mathcal{C}(\rho_f)$ reaching up to 0.953 and $I_2$ up to 2.8281.

By evaluating the eigenvalues and eigenvectors of the arithmetic average of the density matrix $\rho_f$ calculated in the lab frame in this region, we find that the resulting mixed state can be approximately represented as 1\% $(RL + LR)/\sqrt{2}$, 1\% $(RL - LR)/\sqrt{2}$, 7\% $(RR - LL)/\sqrt{2}$, and 90\% $(RR + LL)/\sqrt{2}$ in the lab frame. This specific state explains the near attainment of the upper limit of concurrence and the optimal CHSH value in quantum mechanics, and significantly simplifies the measurement and utilization of the resulting products. In the next section, we will focus on this region as a case study and propose an innovative approach for measuring the polarization correlation implied by this state.

\section{Measurement of the free-traveling entangled electron-positron pairs}

Measurements of electron spin trace back to the Stern-Gerlach experiment published in 1922~\cite{Gerlach:1922dv}, which essentially measures the spin projection of valence electrons in silver atoms along the gradient direction of the magnetic field. However, measuring the spin of a single traveling electron presents significant challenges. For many years following the Stern-Gerlach experiment, it was deemed impossible to spatially separate traveling electrons with different spins due to the comparable effects of the Lorentz force and the uncertainty principle on spin-induced motion~\cite{Batelaan:1997pv}.  It wasn't until 1997 that Stern-Gerlach-like apparatuses, similar to those proposed by Brillouin in 1928~\cite{brillouin1928possible}, were first analytically and numerically demonstrated to achieve this separation using strong magnetic fields~\cite{Batelaan:1997pv,PhysRevLett.81.4772,PhysRevLett.81.4773,PhysRevA.60.63}. Since then, various proposals have emerged to tackle this challenge, primarily focusing on minimizing the Lorentz force in novel Stern-Gerlach-like setups~\cite{mcgregor2011transverse,kohda2012spin,Dellweg:2016njh,Ghazaryan2020-eq}. Most of these proposals target low-energy electrons and may not be effective for relativistic electrons. To date, measuring the spin of flying electrons remains a significant and fundamental research area. Despite the considerable challenges associated with measuring the spin of a single traveling electron, particularly at the GeV scale, the polarization of keV- to GeV-scale electron beams can be measured with a precision of 2–3\% through collective scattering behaviors of beam electrons using various methods~\cite{Aulenbacher:2018weg}.

\begin{figure}
\includegraphics[width=.63\linewidth]{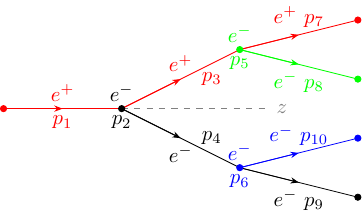}
\caption{\justifying Schematic diagram of the proposed cascade on-target experiment for measuring polarization correlations of the intermediate products. The circles depict the initial and final positions of the particles, while the arrows indicate their momenta. Assuming a 1 GeV incident positron beam with a flux of $10^{10\,(12)}/\mathrm{s}$, the simultaneous production rates of four-body final states can reach the order of $1\,(100)/\mathrm{s}$.}
\label{fig:ttt}
\end{figure}

Building on the fundamental concept of measuring beam polarization correlations through polarized scattering processes, we propose a novel experimental scheme illustrated in Fig. \ref{fig:ttt}. Entangled primary positron and electron beams (Particles 3 and 4) interact with electrons in two separate polarized secondary targets (Particles 5 and 6), carrying polarization information into the angular distributions of the resulting products (Particles 7--10). The electric charges of these products can be measured to distinguish the two occurred polarized processes. To minimize the number of kinematic variables involved, we measure the joint angular distribution of the two secondary scattering processes in their respective center-of-mass frames. Through this experimental setup, the previously concealed polarization correlation of the primary electron-positron pairs is transformed into the joint distribution of the two angular variables in the secondary scattering center-of-mass frames, $\theta^\prime_7$ and $\theta^\prime_9$, which can be derived from the lab-frame angular variables measured by various detectors.

\begin{figure}
\includegraphics[width=.5\linewidth]{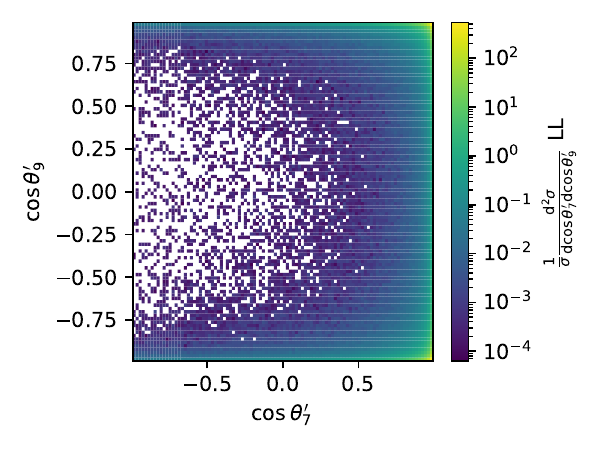}\includegraphics[width=.5\linewidth]{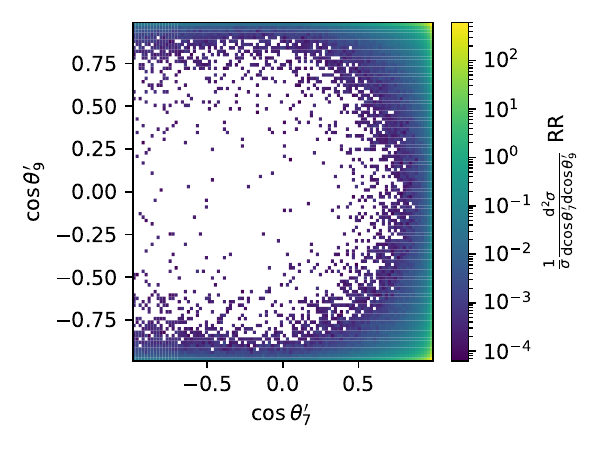}
\includegraphics[width=.5\linewidth]{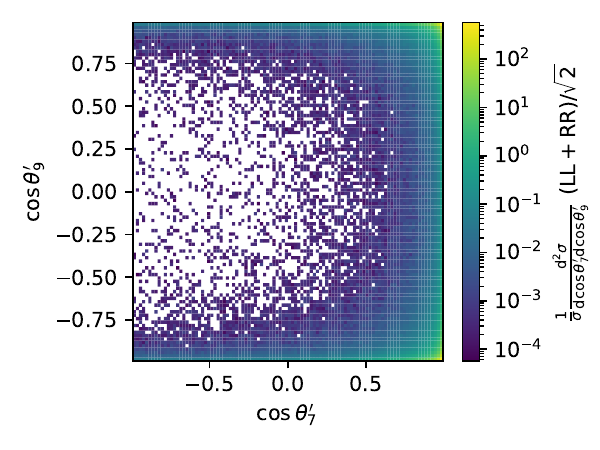}\includegraphics[width=.5\linewidth]{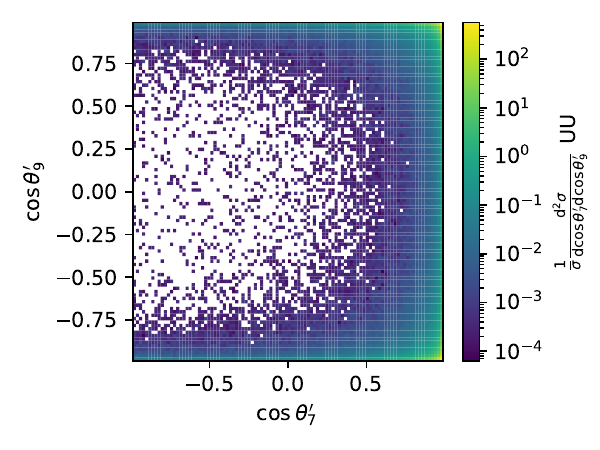}
\caption{\justifying Joint angular distributions of the two secondary scattering processes for four typical states of $s_3s_4$. The events are simulated within the range $0.05\ \mathrm{rad} \leq \theta_3 \leq 0.1\ \mathrm{rad}$ for a 1 GeV positron on-target experiment and are normalized to display the density.}
\label{fig:polar}
\end{figure}

To illustrate the sensitivity of the correlation measurement, we calculate the normalized differential cross section $\frac{1}{\sigma}\frac{\mathrm{d}^2\sigma}{\mathrm{d}\cos\theta^\prime_{7}\mathrm{d}\cos\theta^\prime_{9}}$ using the WHIZARD Monte Carlo event generator~\cite{Kilian:2007gr,Moretti:2001zz}. The incident positron energy is set to 1 GeV, and the primary scattering phase space is restricted to $0.05\ \mathrm{rad} \leq \theta_3 \leq 0.1\ \mathrm{rad}$ as elaborated in the previous section. Given the narrow angular widths of the secondary incident beams, we first consider a simplified ideal scenario where the spins of the electrons in the secondary targets (Particles 5 and 6) are always aligned with the beam direction. This setup maximize the sensitivity to the z-component of the beam polarization. We divide the range of $\theta_3$ into 100 bins and generate $10^5$ secondary scattering events for each bin, using the midpoint value of $\theta_3$ as input. The resulting joint distributions of 4 typical polarization states of Particles 3 and 4 are presented in Fig.~\ref{fig:polar}, where $U$ denotes the unpolarized state. The events are weighted by the product of the differential cross sections of the three scattering processes and normalized to display the density. The state $(LL + RR)/\sqrt{2}$ behaves approximately as the mixed state produced by the primary scattering whose components are given in the previous section. Significant discrimination is observed among the $LL$, $RR$, and $UU$ states, while the distinction between $(LL + RR)/\sqrt{2}$ and $UU$ is relatively weak.

In addition to an incident positron flux of $10^{12}/\mathrm{s}$ and a primary target of $10\ \mathrm{cm}$ thick aluminum (providing Particle 2 in Fig.~\ref{fig:ttt}), we also assume that the two secondary targets are $10\ \mathrm{cm}$ thick iron (providing Particles 5 and 6 in Fig.~\ref{fig:ttt}). The resulting event rate in the region where $\cos\theta^\prime_7 \leq 0.5$ and $-0.75 \leq \theta^\prime_9 \leq 0.75$ is $1.4 \times 10^2$ events per second for the state $\left(LL + RR\right) / \sqrt{2}$. To quantify the difference between the angular distributions resulting from the $\left(LL + RR\right) / \sqrt{2}$ and $UU$ states, we optimize the ratio of the yields of $\left(LL + RR\right) / \sqrt{2}$ to $UU$ in a contiguous smooth area, yielding a value of $1.29 \pm 0.03$, where the uncertainty is estimated with the Poisson counting error for the Monte Carlo events. It corresponds to $4.4 \times 10^3$ post-optimization efficient signal event counts and an expected signal yield over a {\bf 27-second} run, indicating that other uncertainties, such as those from process modeling and background suppression, may dominate the real experimental analysis. The same procedure is applied to the state $\left(LR + RL\right) / \sqrt{2}$ for comparison, resulting in a ratio of $0.78 \pm 0.02$.

For the 20\% polarized targets, the ratios are $1.010 \pm 0.009$ and $0.986 \pm 0.009$ generated from 25 times the number of Monte Carlo events of the 100\% polarized targets to reduce counting errors, with resolving power still present. For the $\left(LL + RR\right) / \sqrt{2}$ signal, it corresponds to $2.5 \times 10^4$ efficient event counts accumulated in {\bf 680 seconds}. The high event rate holds, indicating a promising potential to mitigate the decline in resolving power associated with low target polarization purities in real-world applications.

Despite the representability of the $\theta^\prime_7$-$\theta^\prime_9$ correlation to the $s_3$-$s_4$ correlation and the promising potential event rates for the cascade scattering experiment, reconstructing the density matrix $\rho_{s^\prime_3s^\prime_4s^{\phantom{\prime}}_3s^{\phantom{\prime}}_4}$ from the secondary scattering products presents a significant challenge. A $4 \times 4$ density matrix is typically defined by 15 independent real parameters, which is considerably more than the four pure states we have examined. For arbitrary mixed states, we find significant degeneracy among them which complicates the quantitative reconstruction of the density matrix. However, the correlation effects can still be investigated, and a simplified state tomography can be performed assuming prior knowledge from the primary scattering, which indicates that approximately only a few linear combinations of the bases $RR$, $RL$, $LR$, and $LL$ exist. We encourage the community to further explore the potential of this innovative approach. We believe that establishing a viable method for performing direct quantum state tomography on high-energy scattering products would represent a significant advancement in the field of quantum science.

\section{Summary and Outlook}

In a GeV-order positron on-target experiment, highly entangled electron-positron pairs can be efficiently produced through Bhabha scattering. By filtering events based on specific scattering angles, it is possible to select events with particular levels of entanglement and energies for each entangled particle. This method enables the creation of a controllable source of entangled electron-positron pairs. Simulations with incident positron energies of 1, 3, and 10 GeV yield concurrence values close to one, along with optimal CHSH values near $2\sqrt{2}$ under extreme conditions. These indicators remain promising across a wide range of the lab frame phase space suitable for electron-positron separation and measurements. Although measuring the spin state of a free-traveling electron or positron presents a significant challenge, it is feasible to measure the average polarizations of electron and positron beams. We found that these strategies are generalizable for measuring the polarization correlations of the electron-positron pairs produced by positron targeting through secondary polarized scatterings. Similar studies can also be conducted in electron on-target experiments.
%The efficiency of these measurements is supported by the high intensity of the entangled pair source, which relies on the strong intensity of the primary positron beam that can be provided by future facilities like the STCF in China.

% can produce spin-entangled electrons, paving the way for future quantum teleportation
% https://link.springer.com/article/10.1007/s11128-021-03117-w
% chained experiment

\section*{Acknowledgments}

This work is supported in part by the National Natural Science Foundation of China under Grants No. 12325504, and No. 12061141002.

% The \nocite command causes all entries in a bibliography to be printed out
% whether or not they are actually referenced in the text. This is appropriate
% for the sample file to show the different styles of references, but authors
% most likely will not want to use it.
%\nocite{*}

\bibliography{main}% Produces the bibliography via BibTeX.

\end{document}